\documentclass[twocolumn,showpacs,floatfix,superscriptaddress,amsmath,amssymb,prl]{revtex4}
\usepackage{mathrsfs}
\usepackage{txfonts}
\usepackage{amssymb}
\usepackage{graphicx}
\usepackage{hyperref}
\usepackage{ulem}
 \usepackage{overpic}
 \usepackage{psfrag}
\usepackage{tabularx}
\usepackage{array}
\usepackage{placeins}

\begin{document}

\title{Bond-Alternation-Induced Topological Quantum Gaussian Transition and Topological Quantum Crossover
      in Ising Chains with the Dzyaloshinskii-Moriya Interaction}

\author{Hai Tao Wang}
\affiliation{Centre for Modern Physics and Department of Physics,
Chongqing University, Chongqing 400044, The People's Republic of
China}
 \author{Sam Young Cho}
 \email{sycho@cqu.edu.cn}
\affiliation{Centre for Modern Physics and Department of Physics,
Chongqing University, Chongqing 400044, The People's Republic of
China}

\begin{abstract}
 Non-local orders, entanglement entropy, and quantum fidelity are investigated
 in an infinite-size bond-alternating Ising chain with the Dzyaloshinskii-Moriya
 interaction
 by employing the infinite matrix product state representation
 with the infinite time evolving block decimation method.
 Directly computing two distinct types of finite
 string correlations for very large lattice distances, in contrast to
 an extrapolated extreme value for finite size chains,
 reveals two topologically ordered phases.
 As the bond alternation varies,
 a topological quantum phase transition with continuously variable critical exponents along
 the phase boundary
 occurs between the two Haldane phases
 for the Dzyaloshinskii-Moriya
 interaction stronger than the Ising interaction, while
 a topological quantum crossover between them happens through an intermediate antiferromagentic phase
 demonstrated with the quantum fidelity
 for the Dzyaloshinskii-Moriya
 interaction weaker than the Ising interaction.
 The critical exponents of the order parameters and the central charges from the entanglement entropy
 quantify the universality classes of the phase transition points.
 Anisotropic Heisenberg types of spin chains with bond alternations are finally discussed to share the same criticality.

\end{abstract}

\maketitle

 {\it Introduction.}$-$
 Topologically ordered states \cite{Wen04} beyond
 the Landau paradigm of spontaneous symmetry breaking~\cite{Sachdev}
 have been studied intensively and extensively in condensed matter systems.
 Moreover, such robust states against troublesome decoherence
 are of rapidly growing interest
 in the field of quantum information processing and computation \cite{Nayak,Kitaev}.
 Similarly to the quantum Hall states \cite{Wen04},
 a quantum phase transition without explicit symmetry breaking
 can occur between such topologically ordered phases in spin lattice systems,
 which can be called topological quantum phase transition (TQPT). %
 Examples include
 a spin-$1/2$ model on square lattice \cite{Wen03},
 Kitaev spin-$1/2$ model \cite{Kitaev,Feng}, toric-code model \cite{Vidal09},
 bond-alternating Heisenberg chain \cite{Wang1}, and so on.
 Characterizations of such topological quantum phase transitions have
 become one of the most important topics in quantum many-body systems.

 As a trigger,
 a bond alternation on spin-$1/2$ lattices \cite{Affleck,Almeida,Gibson}
 can generate non-local string orders \cite{Nijs,Tasaki} that can characterize
 a topologically ordered phase, for instance, the Haldane phase \cite{Haldane}.
 In this work, we characterize quantum phase transitions induced by a bond alternation on
 Ising chains with the Dzyaloshinskii-Moriya (DM) interaction.
 By employing the infinite matrix product state (iMPS)
 representation
 with the infinite time evolving block decimation (iTEBD) method
 developed by Vidal~\cite{Vidal},
 we calculate long-range string orders directly for very large lattice distances~\cite{Su}.
 The bond alternation is shown to enable to realize
 the topologically distinct ordered phases distinguished
 by two long-range string orders, respectively.
 Between them, we observe
 (i) a direct phase transition that belongs to the Gaussian universality class,
 or (ii) an indirect transition (crossover) with two Ising-type phase transitions.
 Entanglement entropy  \cite{Cardy,Tagliacozzo}
 verifies
 the phase transitions and their universality classes with
 central charges.
 Groundstate fidelity per lattice site (FLS) \cite{Zhao,Su2,Zhou08}
 also demonstrates groundstate degeneracies that determine a $Z_2$ broken-symmetry phase for the topological quantum crossover (TQC) as well as the phase transitions.

 {\it Model and string order parameters.}$-$
 Let us consider the model Hamiltonian
\begin{equation}
 H = \sum_{j=-\infty}^{\infty} \Big(1+(-1)^{j}\, \delta\Big)
  \Big[  \mbox{\boldmath $D$} \cdot
  \mbox{\boldmath $S$}_{j} \times \mbox{\boldmath $S$}_{j+1}
 +  J \, S^z_{j}S^z_{j+1} \Big],
 \label{HamDM}
\end{equation}
 where $S^{\alpha}_i$ $(\alpha \in x,y,z)$ are the spin-1/2 operators at
 lattice site $i$, $J (> 0)$ and $\mbox{\boldmath $D$}$ denote the Ising
 and the DM interactions,
 respectively.
 The bond-alternation parameter ranges as $-1 \leq \delta \leq 1$.
 We choose $\mbox{\boldmath $D$}=D\, \hat{z}$ with $D > 0$.
 For clarity, we study mainly the system in Eq. (\ref{HamDM}).
 However, adding a term
$\sum_j \Big(1+(-1)^{j}\, \delta\Big) (\, S^x_{j}S^x_{j+1} +  S^y_{j}S^y_{j+1})$
 in Eq. (\ref{HamDM}) does not change the physics of criticality,
 will be discussed later.

\begin{figure}
\includegraphics [width=0.45\textwidth]{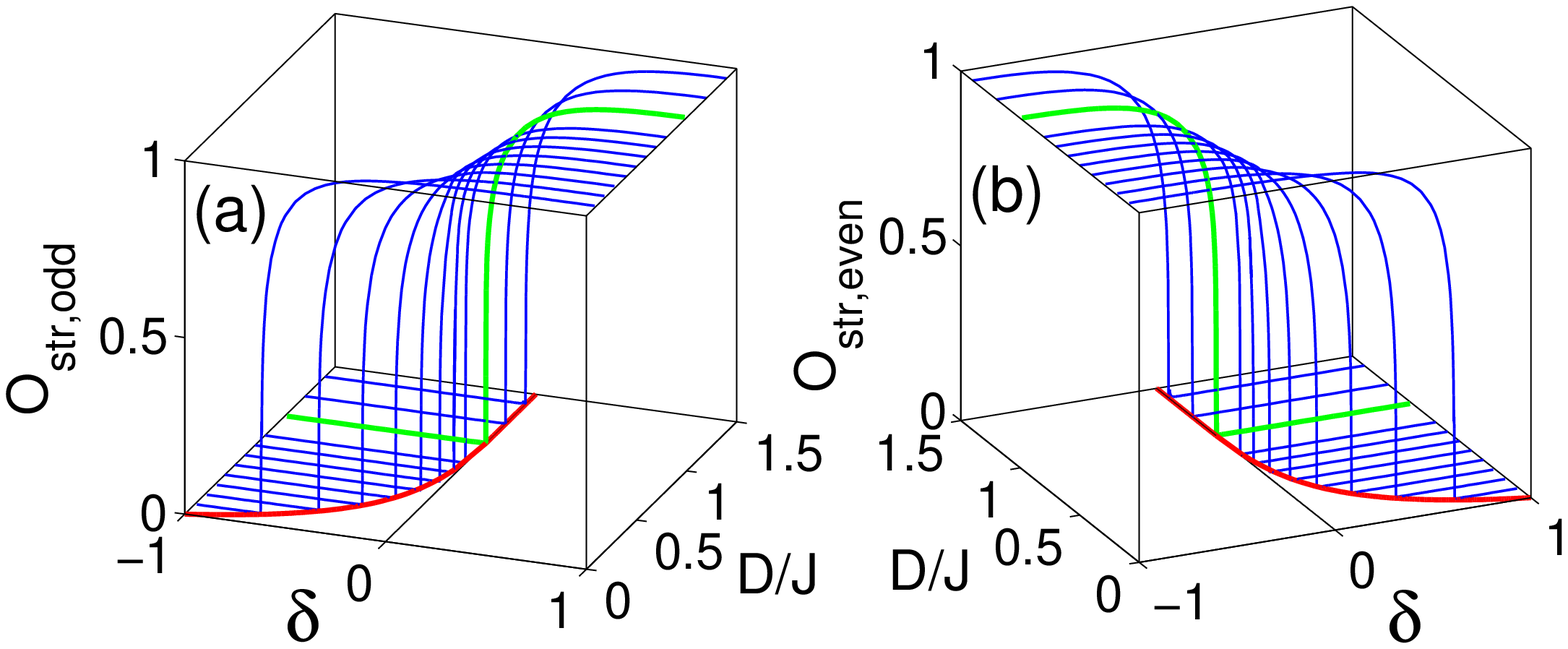}
\includegraphics [width=0.45\textwidth]{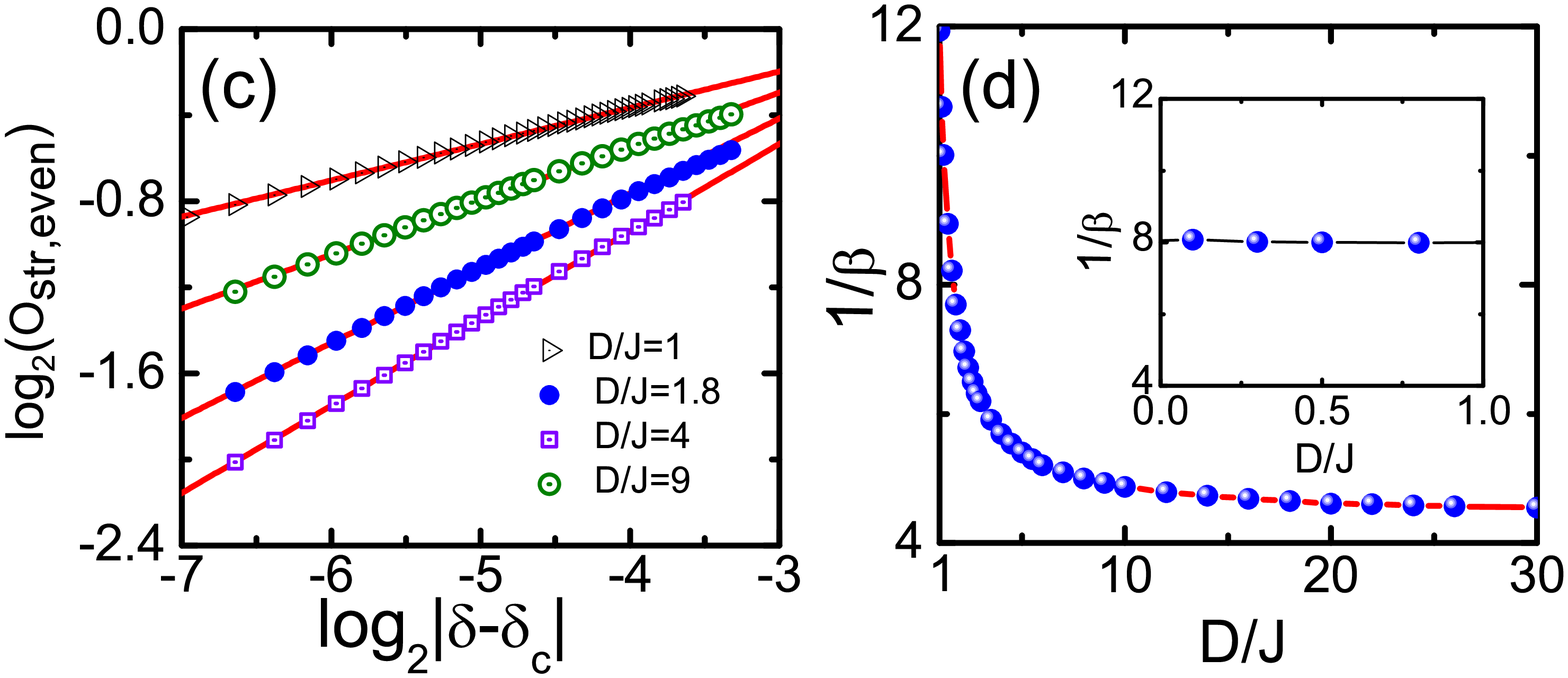}
\caption{ (Color online) (a) Even and (b) odd string order
       parameters, $O_{str,even/odd}$, in the $\delta$-$(D/J)$ plane.
 (c)
  $O_{str,even}$ as a function of $|\delta-\delta_c|$
  for various DM interactions
  $D=J$, $1.8J$, $4J$, and $9J$, corresponding to
  the numerical critical exponents $\beta=1/12$, $1/8$,
  $1/5.688$, and $1/5$, respectively.
  (d) $\beta$ as a function of $D/J$ in the $O_{str,even}$.
 Here, note that $O_{str,even}(\delta)=O_{str,even}(-\delta)$.
 In (a) and (b), the red lines are the phase boundaries.
 The truncation dimension is $\chi=32$.
  }
  \label{fig1}
\end{figure}


  Based on the bond
  alternation, one can define two string order parameters ~\cite{Hida1,Hyman} as
%
 \begin{subequations}
 \begin{eqnarray}
 O^{\, \alpha}_{str,even}  \!\!\! &=&  \!\!\!  \lim_{|i-j|\rightarrow\infty}
 \left(-4\left\langle
 S^\alpha_{2i} \exp \left[i\pi\sum_{k=2i+1}^{2j-2}S^\alpha_k\right]
         S^\alpha_{2j-1}\right\rangle \right), \\
 O^{\, \alpha}_{str,odd}  \!\!\!  &=&  \!\!\!
 \lim_{|i-j|\rightarrow\infty} \left(-4\left\langle S^\alpha_{2i+1} \exp
 \left[i\pi\sum_{k=2i+2}^{2j-1}S^\alpha_k \right] S^\alpha_{2j}\right\rangle \right),
 \label{equ:string}
\end{eqnarray}
\end{subequations}
%
 where $\alpha=x$, $y$, and $z$. Our iMPS groundstate
 wavefunction allows us to directly calculate the defined string orders
 \cite{Su}.
 Figure \ref{fig1}(a) and \ref{fig1}(b) show, respectively,
 the even  and the odd string order parameters $O^{\, z}_{str,even/odd}$ in $\delta$-$(D/J)$ plane.
 In fact, $O_{str,even}(\delta)=O_{str,odd}(-\delta)$ for a given $D/J$.
 As the bond alternation $\delta$ varies,
 the even (odd) string order parameter is finite
 for $\delta > \delta^{\,+}_c$ ($\delta < \delta^{\,-}_c$),
 where the phase boundary functions are obtained as
 \begin{eqnarray}
  \delta^{\,\pm}_c (J,D) =
   \mp\, \delta_c\, \Theta(J-D)
  \label{bondaries}
 \end{eqnarray}
 with $\delta_c = (D-J)^2/(A D + J)^2$, the numerical fitting constant $A \simeq 5/4$,
 and
 the unit step function $\Theta (x)$ representing equal to $0$ for $x < 0$
 and $1$ for $x \geq 0$.
  Other components of the string order parameters are zero, i.e.,
  $O^{\, x/y}_{str,even/odd} = 0$.
  Thus, the finite even (odd) string order parameter characterizes
  a topologically ordered
  phase, i.e., the gapful even (odd) Haldane phase~\cite{Nijs,Tasaki,Oshikawa}.
 Similarly to the spin-1 Heisenberg chain understood by
 the hidden $Z_2\times Z_2$ breaking symmetry~\cite{Kennedy},
 a similar hidden symmetry breaking may occur for each phase.

  In addition, the phase boundaries of topological characterizations in Eq. (\ref{bondaries})
  expose two ways changing from one Haldane
  phase to the other Haldane phase as the bond
  alternation varies.
  (i) For $J \leq D$,
  the even (odd) string order parameter is finite
  for $\delta > 0$ ($\delta < 0$) with
  $\delta^{\,\pm}_c=0$, which implies
  that
  a TQPT occurs at $\delta=0$.
  (ii) For $J > D$, i.e., $\delta^{\,\pm}_c=\mp \delta_c$,
  the even and the odd string order parameters have a finite value
  for $\delta > -\delta_c$ and $\delta < \delta_c$, respectively.
  The system is then in the odd (even) Haldane phase for $\delta < -\delta_c$
  ($\delta > \delta_c$).
  However,
  the order parameters coexist for $-\delta_c < \delta < \delta_c$.
  This implies that
  a TQC occurs
  from the even Haldane phase to the odd Haldane phase or vice versa
  in the range of the bond alteration, $-\delta_c < \delta < \delta_c$.
 In some sense, our system with such topological phase changes resembles
 an anisotropic antiferromagnet,
 possessing a magnetic crossover with
 two continuous phase transitions at phase boundaries,
 such as GdAlO$_3$,
 where anisotropies favors spin alignment along particular lattice directions,
 breaks an $O_n$ symmetry,
 and give rise to a multicritical point, particularly, a tetracritical point
  \cite{Chaikin,Fisher,Rohrer}.

 {\it Topological quantum Gaussian transition.}$-$
 In order to obtain the critical exponents of the string order parameters,
 in Fig. \ref{fig1}(c), we plot the string order parameter $O_{str,even}$ as a
 function of $|\delta-\delta^{\, +}_c|$ for various values
 of $D/J$.
 Note that the
 string order parameters scale as $O_{str,even} \propto |\delta-\delta^{\, \pm}_c|^{2\beta}$
 and the critical exponents $\beta(D/J)$ depend on the chosen values of $D/J$.
 Figure \ref{fig1}(d) shows the $(D/J)$ dependence of the critical exponents.
 For the TQPT with $J \leq D$,
 the critical exponent $\beta(D/J)$ increases from $\beta(1)=1/12$
 as $D/J$ increases from $D/J=1$.
 Especially, for $J=D$ [Fig. \ref{fig1}(c)], the critical  exponent $\beta =1/12$
 corresponds to the value in the bond-alternating spin-$1/2$
 Heisenberg chain \cite{Hida1,Wang1}.
 Also, for $D=4J$ [Fig. \ref{fig1}(c)],
 the critical  exponent $\beta =1/5.688$
 is very close to the value $\beta =1/\sqrt{32}$
 predicted for the surface-roughening transition
 in the 2D classical model \cite{Luther}
 and the Gaussian transition in
 the spin-$1$ chain with single-ion anisotropy \cite{Hu}.
 Such correspondences of the critical exponents in different physical systems
 demonstrates the common physics of the Gaussian type transitions.
 Furthermore, by using the critical exponent
 function $1/\beta(\lambda) = (B_1/\pi) \cos^{-1} [-\lambda]-B_2$
 for the Gaussian transition in the Ashkin-Teller model ($B_1=16$ and $B_2=4$) \cite{Kohmoto},
 our numerical fitting denoted by the solid line gives
 the numerical constants $B_1=16.68$ and $B_2=3.95$ with $\lambda = J/D$ [Fig. \ref{fig1}(d)].
 The continuous variable critical exponent shows that the TQPT for $J \leq D$
 is a {\it topological quantum Gaussian transition}, while
 the inset of Fig. \ref{fig1}(d) shows
 an Ising type of continuous phase transitions for $J > D$ because
 $\beta =1/8$ at $\delta = \pm \delta_c$.
%
%

\begin{figure}
\includegraphics [width=0.45\textwidth]{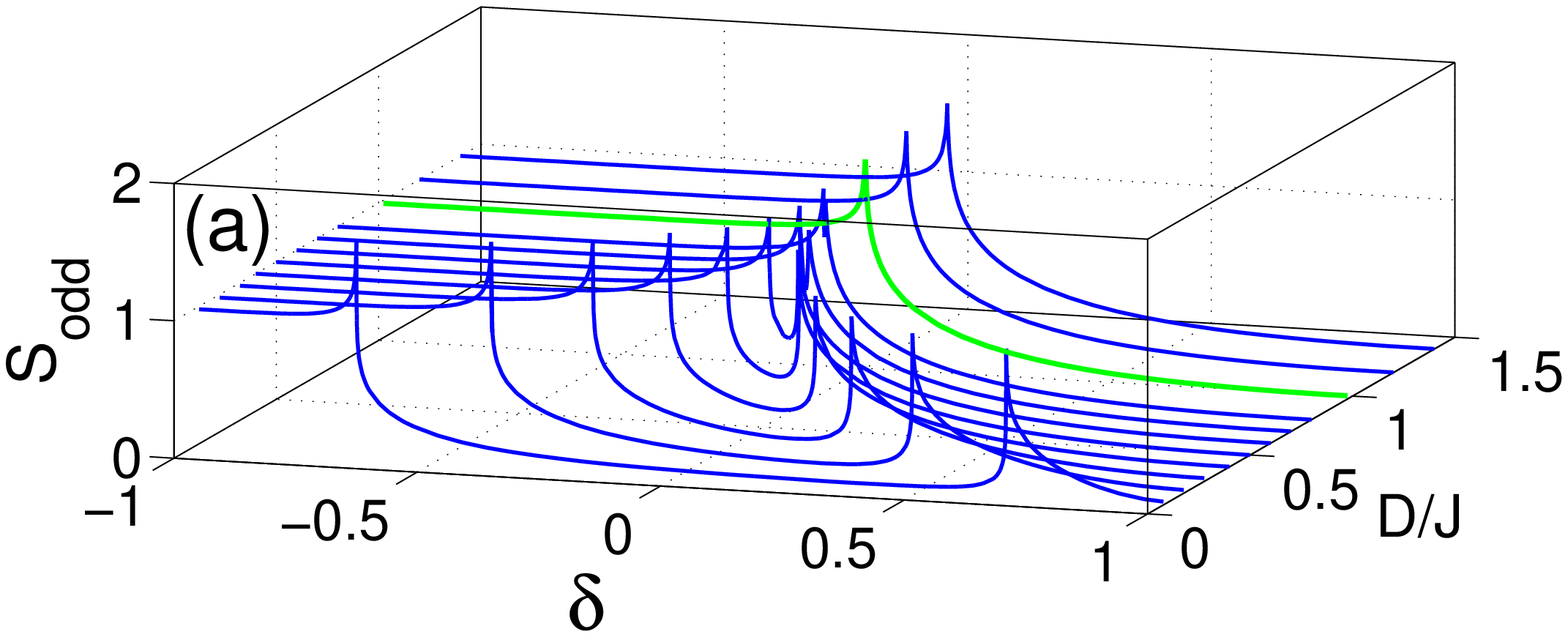}
\includegraphics [width=0.5\textwidth]{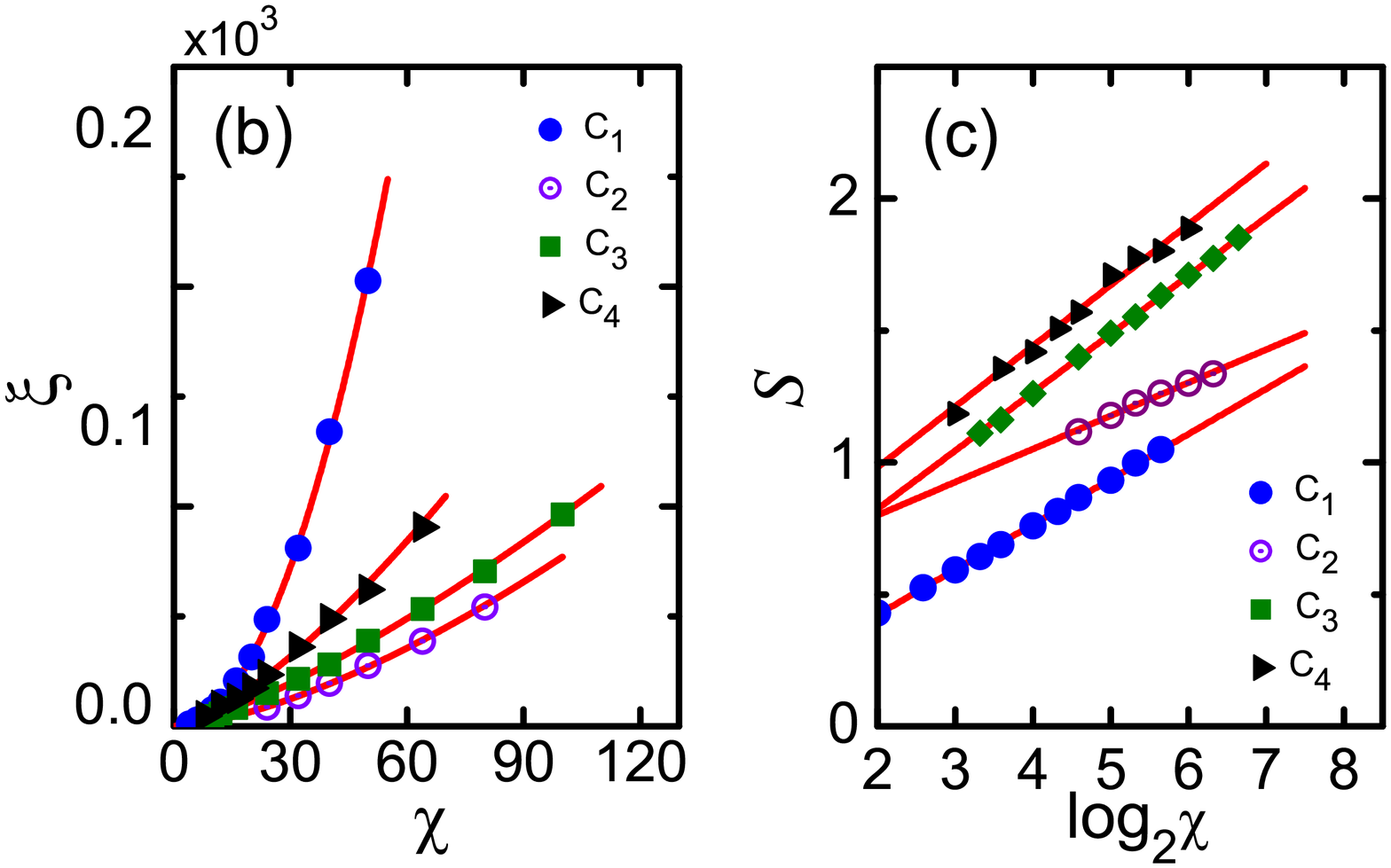}
\caption{ (Color online)
  (a)  Von Neumann entropies
  $S_{odd}$ in the $\delta-(D/J)$ plane with $\chi=32$.
  Here, $S_{odd}(\delta)=S_{even}(-\delta)$.
  (b) Divergence of the correlation lengths $\xi$ as a function of the
  truncation dimension $\chi$ at the critical points.
   (c) Divergence of von Neumann entropies as a function of  $\chi$ at the critical points in (b).
   The C's are given in the text.
  }
  \label{fig2}
\end{figure}

 {\it Quantum entanglements and central charges.}$-$
 Singular behaviors of the von Neumann entropy reveal quantum critical behavior.
 They have been verified to obey a universal
 scaling law in one-dimensional lattice systems in the thermodynamic
 limit
 and, at critical points, are related to a universal factor, i.e., a central
 charge of associated conformal field theory \cite{Cardy,Tagliacozzo}.
 Let us consider quantum entanglements between two half-infinite subsystems.
 Our system is partitioned into
 the left and the right half-infinite subsystems denoted by $L$ and $R$.
 The von Neumann entropy between $L$ and $R$ is defined
 as $S = -\mathrm{Tr} \rho_{L}\log_{2}\rho_{L}
 =-\mathrm{Tr} \rho_{R}\log_{2}\rho_{R}$ in terms of
 the reduced density matrix of subsystems $\rho_{L}$ or $\rho_{R}$.
 In the iMPS representation,
 the von Neumann entropy can be expressed in terms of the Schmidt coefficients, $\lambda_{\alpha}$, as
 $
 S = -\sum_{\alpha=1}^{\chi} \lambda^{2}_{\alpha}\log_{2}\lambda^{2}_{\alpha}.
 $
 Due to the bond alternation, there are two types of Schmidt coefficient
 matrices that describe two possible ways
 of the partitions, i.e., one is on the even sites, the other is on
 the odd sites.

 In Fig.~\ref{fig2}(a), the von Neumann entropy $S_{odd}$
 is plotted in $\delta$-$(D/J)$ plane.
 The singular behaviors (peaks) of the entropy indicate the phase
 transition along the phase boundaries $\delta=\delta^{\, \pm}_c$.
 Note that the entropy peak at $\delta=0$ for $J \leq D$ is split into the two peaks
 at $\delta = \pm \delta_c$ for $J > D$.
 Since the two von Neumann entropies
 $S_{even/odd}$ depending on the
 even- or the odd-site partitions satisfy $S_{odd}(\delta)=S_{even}(-\delta)$,
 the $S_{odd}$ has the same singular behaviors along the phase boundaries $\delta=\delta^{\, \pm}_c$.
 In Fig.~\ref{fig2}(b), the correlation lengths $\xi(\chi)$ are plotted as a function of
 the truncation dimension $\chi$ for the critical points.
 As the truncation dimension increases,
 the correlation lengths $\xi$ scale to diverge as
 $\xi (\chi) = \xi_0 \, \chi^\kappa$,
 which means the scale invariance of the system
 in the thermodynamic limit \cite{Cardy,Tagliacozzo}, with the
 numerical finite-entanglement scaling exponents $\kappa$,
  (i) $\xi_0=0.073$ and $\kappa=2.031$ at $C_1(\delta,D/J)=(0.2651,0.3)$,
  (ii) $\xi_0=0.0727$ and $\kappa=1.5126$ at $C_2=(0.0085,0.8)$,
  (iii) $\xi_0=0.217$ and $\kappa=1.323$ at $C_3=(0,1)$,
  and  (iv) $\xi_0=0.298$ and $\kappa=1.376$ at  $C_4=(0,4)$.
 In Fig.~\ref{fig2}(c), we show the logarithmic scaling of the von Neumann entropy $S(\chi)$
 for the critical points.
 From the $\kappa$'s in Fig. \ref{fig2}(b), the linear fittings
   $S(\chi)=S_0 + (c\kappa/6)\log_2{\chi}$ \cite{Tagliacozzo} yield
   (i) the central charge $c\approx 0.505$ with $S_0=0.0809$ at
   $C_1$,
   (ii) $c\approx 0.4979$ with $S_0=0.5491$ at $C_2$,
   (iii) $c\approx 1.002$ with $S_0=0.3816$ at $C_3$,and
    (iv) $c\approx 1.003$ with $S_0=0.521$ at $C_4$.
 Consequently, the characteristic entanglement properties
 confirm that for $J \leq D$,
 the TQPT between the even- and the
 odd-Haldane phases at the critical points $\delta=0$
 is a Gaussian transition which is characterized
 by the central charge $c=1$
 and the occurrence of a phase transition between two gapful phases
 with the continuous variable critical exponent of the string order
 parameters.
 For $J > D$, the central charge $c=1/2$ at the
 phase boundaries $\delta=\pm \delta_c$ also confirms that an Ising type of
 phase transitions occurs along the boundaries of the TQC.

\begin{figure}
\includegraphics [width=0.5\textwidth]{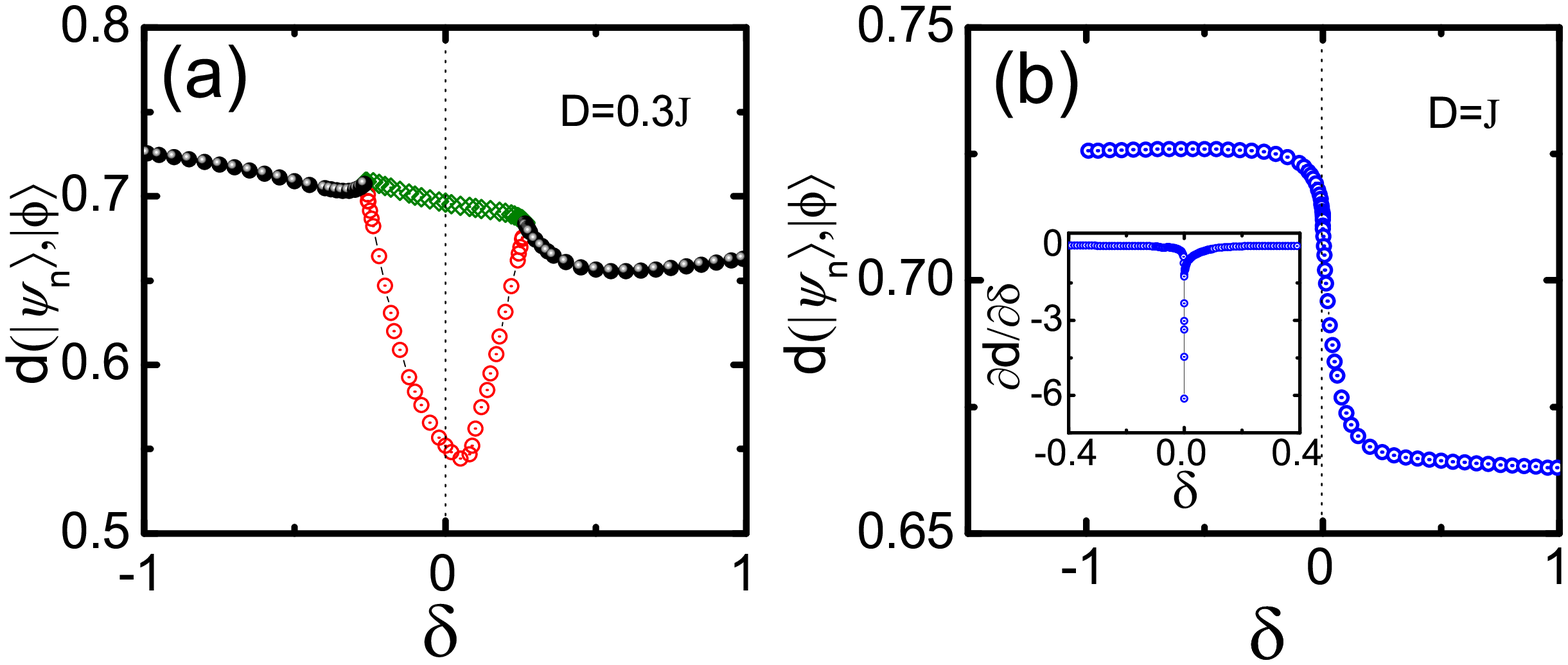}
\includegraphics [width=0.5\textwidth]{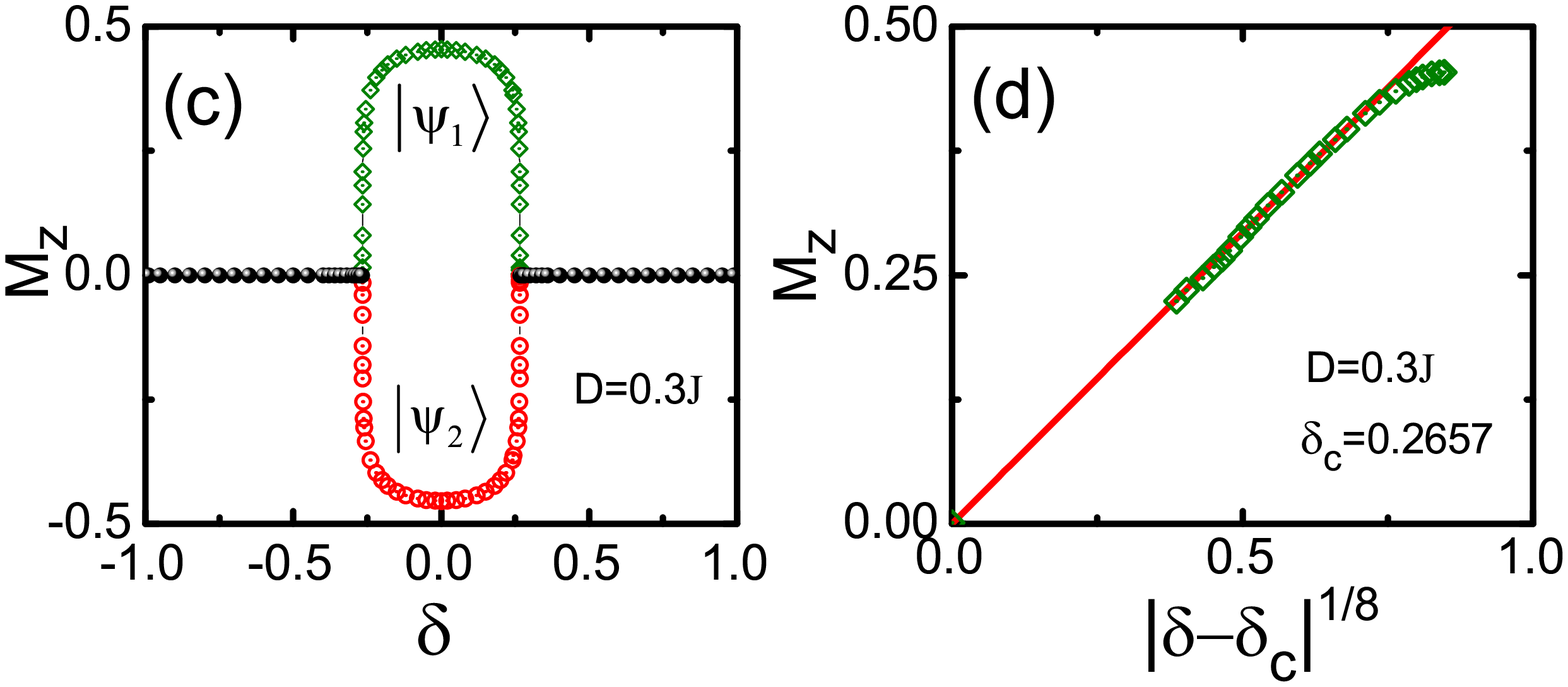}
\caption{ (Color online)
 Fidelity per site $d(|\psi_n\rangle,|\phi\rangle)$
 as a function $\delta$ for (a) $D = 0.3J$ and (b) $D=J$
 (${\partial d}/{\partial \delta}$ in the inset).
 (c) Staggered magnetizations $M_z$ from the two groundstates
 $|\psi_1\rangle$ and $|\psi_2\rangle$ for $D=0.3J$.
 (d) The critical exponent $\beta=1/8$ from $M_z \propto |\delta-\delta_c|^{1/8}$
 for $D=0.3J$ and $\delta_c = 0.2657$.
}
  \label{fig3}
\end{figure}

 {\it Quantum fidelity per site and intermediate antiferromagnetic phase
 for topological quantum crossover.}$-$
 Although the TQPT occurring without any explicit symmetry breaking
 has been understood by the string order parameters,
 our system can undergo a symmetry breaking
 because
 the Hamiltonian in Eq. (\ref{HamDM}) is invariant
 under the unitary transformation
 $U = \prod U_{2i}\otimes U_{2i+1}$
 with $U_{2j}=\sigma^x$ and 
 $U_{2j+1}=\sigma^y$, i.e., 
 $UHU^\dagger=H$, and then
 possesses a $Z_2$ symmetry
 generated by the transformation $U$.
 If the system undergoes explicitly a spontaneous breaking
 of the $Z_2$ symmetry in the interaction parameter space,
 it has a $Z_2$ broken-symmetry phase with a doubly degenerate groundstate.
 Thus, in order to clarify whether the $Z_2$ symmetry breaking occurs,
 let us consider a quantum fidelity that allows us to determine
 groundstate degeneracy in one-dimensional infinite quantum lattice
 systems \cite{Zhao,Su2}.
 We employ the FLS $d(|\psi_{n}\rangle,|\phi\rangle)$ in
 Ref. \onlinecite{Su2} as
$ \ln d(|\psi_{n}\rangle,|\phi\rangle) \equiv \lim_{L \rightarrow
\infty}
  (1/L) \ln F(|\psi_{n}\rangle,|\phi\rangle)$,
 where the quantum fidelity is $F(|\psi_{n}\rangle,|\phi\rangle)=|\langle \psi_{n}|\phi\rangle|$,
 $L$ is the system size, $|\psi_{n}\rangle$ is an iMPS groundstate calculated with
 the randomly chosen $n$-th initial state for given parameters, and $|\phi\rangle$ is an arbitrary chosen reference state. If $F$ has $N$ projection values onto the reference state, the system
 has $N$ degenerate groundstates.
 By using many random initial states for given parameters,
 we detect a degenerate groundstate.
 In Fig. \ref{fig3},
 the FLS $d(|\psi_{n}\rangle,|\phi\rangle)$ shows
 that the system has a doubly generate groundstate for $-\delta_c < \delta < \delta_c$ and $J > D$
  [Fig. \ref{fig3}(a)]
 and a single groundstate for $J \leq D$ [Fig. \ref{fig3}(b)].
 Note that the bifurcation points \cite{Zhao,Su2} at $\delta=\pm \delta_c$
 correspond to the phase boundaries [Fig. \ref{fig3}(a)]
 and the singular behavior \cite{Zhou08,Rams} at $\delta=0$ in the derivative of the FLS over the
 bold alternation $\delta$
 indicates the phase transition point [the inset of Fig. \ref{fig3}(b)].
 As a result,
 the doubly degenerate groundstate
 implies that the topological quantum crossover region, i.e., $-\delta_c < \delta <
 \delta_c$ for $J > D$, is a $Z_2$ broken-symmetry phase.

 For the $Z_2$ broken-symmetry phase, actually,
 there are the two groundstates $\lbrace |\psi_g\rangle, U |\psi_g\rangle\rbrace$
 that satisfy $H |\psi_g\rangle = E_g |\psi_g\rangle$
 or $H U|\psi_g\rangle = E_g U|\psi_g\rangle$ with the groundstate energy $E_g$,
 and they are not equal, i.e., $|\psi_g\rangle \neq U |\psi_g\rangle$.
 One can then denote $|\psi_1\rangle=|\psi_g\rangle$ and $|\psi_2\rangle=U|\psi_g\rangle$.
 Due to the transformations of the spin operators as
  $ U S^z_j U^\dagger  = - S^z_j $
  and
  $ U S^z_j S^z_{k} U^\dagger  =  S^z_j S^z_{k} $,
 the local magnetizations and the spin-spin correlations
 from the two groundstate wavefunctions might have the relations
 $\langle \psi_1|S^z_j |\psi_1\rangle = -\langle \psi_2| S^z_j |\psi_2\rangle$
 and
 $\langle \psi_1| S^z_j S^z_{k} |\psi_1\rangle = \langle \psi_2| S^z_j S^z_{k} |\psi_2\rangle$,
 respectively.
 Further, for $J > D$, the nearest spin-spin correlation is antiferromagnetic,
  i.e., $\langle \psi| S^z_j S^z_{j+1} |\psi\rangle < 0$.
 In Fig. \ref{fig3}(c), then,
 we plot the staggered magnetization $M_z=\langle (S^z_j-S^z_{j+1})/2\rangle $
 as a function of $\delta$.
 The two groundstates give
 a finite staggered magnetization for $-\delta_c < \delta <
 \delta_c$ and
 $M_z=\langle \psi_1| (S^z_j-S^z_{j+1})/2|\psi_1\rangle =
 - \langle \psi_2| (S^z_j-S^z_{j+1})/2|\psi_2\rangle$.
 The string order parameters calculated from the two degenerate groundstates
 are the same each other, i.e.,
 $ \langle \psi_1| O_{str,even/odd}|\psi_1\rangle =\langle \psi_2| O_{str,even/odd}|\psi_2\rangle$,
 as it should.
 In addition, Fig. \ref{fig3}(d) shows that
 the staggered magnetization scales as $M_z \propto
 |\delta-\delta_c|^{\beta}$ with the critical exponent $\beta=1/8$.
 Consequently,
 the TQC region
 is characterized by the local order, i.e., the staggered magnetization.
 The valance bond solid picture \cite{Affleck} may state
 that a nonlocal string order for a Haldane phase captures so-called `dilute'
 antiferromagnetic phase.
 Hence,
 the TQC between the two distinct `dilute' antiferromagnetic
 phases, i.e., the even- and the
 odd-Haldane phases, occurs via the intermediate antiferromagnetic state
 as the bond alternation $\delta$ varies for $ J > D$.

\begin{figure}
\includegraphics [width=0.45\textwidth]{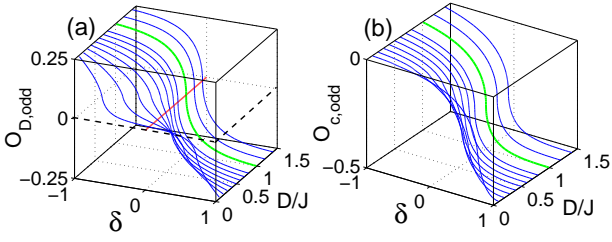}
\caption{ (Color online) (a) Dimer order $O_{D,odd}$ and (b) Chiral order $O_{C,odd}$
    in the $\delta$-$(D/J)$ plane.
%
%
 }
  \label{fig4}
\end{figure}

 {\it Dimer and chiral orders.}$-$
 Due to the symmetry of the Hamiltonian in Eq. (\ref{HamDM}),
 the system can have the other local orders that have nothing to do with the $Z_2$ symmetry breaking.
 To show this point explicitly, in Fig. \ref{fig4}(a) and \ref{fig4}(b),
 the odd dimer and the odd chiral orders are plotted
 in $\delta-(D/J)$ plane.
 Here, the dimer and the chiral orders are defined as
 $ O_{D,even} = \langle\vec S_{2j} \cdot \vec S_{2j+1} -\vec S_{2j+1} \cdot \vec S_{2j+2}\rangle$,
 $ O_{D,odd} = \langle\vec S_{2j-1} \cdot \vec S_{2j} -\vec S_{2j} \cdot \vec
 S_{2j+1}\rangle$,
%
 $ O_{C,even} = \langle \vec S_{2j} \times \vec S_{2j+1}  \rangle_z$,
 and $ O_{C,odd} = \langle \vec S_{2j-1} \times \vec S_{2j}\rangle_z$.
 They satisfy the relations
    $O_{D,odd}(\delta)=O_{D,even}(-\delta)$ with $O_{D}(0)=0$
    and $O_{C,odd}(\delta)=O_{C,even}(-\delta)$ with $O_{C}(0) \neq 0$.
 As they should be,
$ \langle \psi_1| O_{D,even/odd}|\psi_1\rangle =\langle \psi_2| O_{D,even/odd}|\psi_2\rangle$
 and
$ \langle \psi_1| O_{C,even/odd}|\psi_1\rangle =\langle \psi_2| O_{C,even/odd}|\psi_2\rangle$.
 Both the dimer and  the chiral orders are finite
 in the whole parameter range, which implies that they cannot distinguish
 the two Haldane phases.
 Hence, only the two string order parameters
 distinguish the two Haldane phases.

 {\it Relations to other spin models.}$-$
 Actually,
 our results can be shared to understand other spin-$1/2$ lattice models with bond alternations.
 Using
 the non-local transformation
  $\tilde H = \exp[-i\sum_j \alpha_j S_j^z] H
 \exp[i\sum_j \alpha_j S_j^z]$,
  the anisotropic Heisenberg chain with the DM interaction
  $H' = \sum_j \Big(1+(-1)^{j}\, \delta\Big)
  \Big[  \mbox{\boldmath $D$} \cdot
  \mbox{\boldmath $S$}_{j} \times \mbox{\boldmath $S$}_{j+1}
 + \Delta (\, S^x_{j}S^x_{j+1} +  S^y_{j}S^y_{j+1})+ J \, S^z_{j}S^z_{j+1} \Big]
 $ can be mapped to a
 spin-$1/2$ XXZ chain with the bond alternation,
 $\tilde H'
 = \sum_j (1+(-1)^{j}\, \delta)
  [\sqrt{D^2+\Delta^2} ( S^x_{j}S^x_{j+1}+ S^y_{i}S^y_{j+1} ) + J \,
  S^z_{j} S^z_{j+1}]$,
 with
 $\alpha_{j}-\alpha_{j+1} =  \tan^{-1}[D/\Delta]$
 or a form of Eq. (\ref{HamDM}),
  $\tilde H'
 = \sum_j (1+(-1)^{j}\, \delta)
  [\sqrt{D^2+\Delta^2} ( S^x_{j}S^y_{j+1}- S^y_{i}S^x_{j+1} ) + J \,
  S^z_{j} S^z_{j+1}]$,
 with
 $\alpha_{j}-\alpha_{j+1} =  \tan^{-1}[D/\Delta]-\pi/2$.
 Thus, for $\Delta=0$,
 Eq. (\ref{HamDM}) can be mapped to a
 spin-$1/2$ XXZ chain with the bond alternation,
 $\tilde H
 = \sum_j (1+(-1)^{j}\, \delta)
  [ D ( S^x_{j}S^x_{j+1}+ S^y_{i}S^y_{j+1} ) + J \,
  S^z_{j}S^z_{j+1}],
  $ with $\alpha_{j}-\alpha_{j+1} = \pi/2$.
  For $D=0$,
  the $H'$ can also be mapped to
  $\tilde H'
 = \sum_j (1+(-1)^{j}\, \delta)
  [\Delta ( S^x_{j}S^y_{j+1}- S^y_{i}S^x_{j+1} ) + J \,
  S^z_{j} S^z_{j+1}]$
 with
 $\alpha_{j} -\alpha_{j+1} = -\pi/2$.
 Hence,
 the bond alternation
 can lead the even and the odd Haldane phases in the systems described by
 the Hamiltonian $H'$
 with the DM interaction.
 Furthermore, according to Kohmoto, den Nijs, and Kadanoff  \cite{Kohmoto},
 the transformed spin-$1/2$ XXZ chain with the bond alternation can be
 mapped to the one dimensional quantum Ashkin-Teller model,
 $H_{AT}=-\sum (\sigma^z_j \sigma^z_{j+1}+\tau^z_j
 \tau^z_{j+1} + \lambda \sigma^z_j \sigma^z_{j+1}\tau^z_j
 \tau^z_{j+1} ) -\eta \sum (\sigma^x_j + \tau^x_j + \lambda
 \sigma^x_j \tau^x_j )$ with $\lambda = J/\sqrt{D^2+\Delta^2}$ and $\delta = (\eta-1)/(\eta+1)$,
 where $\sigma^\alpha_j$ and $\tau^\alpha_j$ are Pauli matrices.
 This Ashkin-Teller model possesses a $Z_2 \times Z_2$ symmetry because it is invariant
 under the unitary transformation $\sigma^z_j \rightarrow -\sigma^z_j$
 and $\tau^z_j \rightarrow -\tau^z_j$.
 Then, for $\Delta=0$, our topological quantum Gaussian critical line corresponds to
 the $\delta=0$ line for $-1/\sqrt{2} < \lambda(=J/D) < 1$
 and at the $\delta=0$ point of the $\lambda(=J/D)=1$
 the critical line split into two Ising critical lines $\delta=\pm\delta_c$
 for our TQC $\lambda(=J/D) > 1$.
 The continuously varying critical exponents in our topological
 quantum Gaussian transition line [Fig. \ref{fig1}(d)]
 agrees well with the exact critical exponent function
 for the Gaussian critical line of the
 Ashkin-Teller model.
 At the point $\delta=0$ for $\lambda(=D/J)=1$,
 our model and the spin-$1/2$ XXZ chain are $U(1)$ symmetric,
 whereas the Ashkin-Teller model is $Z_2 \times Z_2$ symmetric.
 The university class of the point $\delta=0$ for $D/J=1$
 is of the Berezinskii-Kosterlitz-Thouless type.

 {\it Summary.}$-$
 We have established that the Ising chain with the DM interaction
 is in the even- or the odd Haldane phases induced by the bond alternation.
 For $D \geq J$, the direct transition between the topologically ordered states belongs to
 the Gaussian type transition.
 For $D < J$, the indirect phase transition is undergone through
 the intermediate antiferromagnetic state that is distinguished by the Ising type of quantum phase transitions from the topologically ordered states.
 In addition,
 the quantum entanglement was shown to detect the TQPTs
 as well as to classify the universality classes.
 The FLS was shown to be a useful tool to detect the degenerate groundstates indicating
 a spontaneous broken-symmetry phase as well as the TQPTs.
 Those results have been obtained from the iMPS numerical calculation.
 We have finally discussed that the same critical phenomena can be seen in
 various anisotropic Heisenberg types of spin chain models with the bond alternation.

\acknowledgments
 We thank Huan-Qiang Zhou for useful comments.
 This work was supported by the National Natural Science Foundation of
 China under the Grant No. 11374379.


\begin{thebibliography}{99}

 \bibitem{Wen04}
 For a review, see, e.g., X. G. Wen, \textit{ Quantum Field Theory of Many-body Systems}
 (Oxford University, Oxford, 2003);


 \bibitem{Sachdev}
 S. Sachdev, \textit{ Quantum phase transitions}
 (Cambridge University Press, Cambridge, England, 1999).

 \bibitem{Nayak}
 C. Nayak, S. H. Simon, A. Stern, M. Freedman, and S. Das Sarma,
 Rev. Mod. Phys. \textbf{80}, 1083 (2008).




\bibitem{Kitaev}
 A. Y. Kitaev, Ann. Phys. (NY) \textbf{303}, 2 (2003).




\bibitem{Wen03}
 X. G. Wen,
 Phys. Rev. Lett. \textbf{90}, 016803 (2003).


\bibitem{Feng}
 X.-Y. Feng, G.-M. Zhang, and T. Xiang,
 Phys. Rev. Lett. \textbf{98}, 087204 (2007).



\bibitem{Vidal09}
 J. Vidal, R. Thomale, K. P. Schmidt, and S. Dusuel,
 Phys. Rev. B \textbf{80}, 081104 (2009).


\bibitem{Wang1}
 H. T. Wang, B. Li, and S. Y. Cho,
 Phys. Rev. B \textbf{87}, 054402 (2013).




 \bibitem{Affleck}
 I. Affleck, T. Kennedy, E.H. Lieb, and H. Tasaki,
 Phys. Rev. Lett. \textbf{59}, 799 (1987).


 \bibitem{Almeida}
 J. Almeida, M. A. Martin-Delgado, and G. Sierra,
 Phys. Rev. B \textbf{76}, 184428 (2007).



 \bibitem{Gibson}
 S. J. Gibson, R. Meyer, and G. Y. Chitov,
 Phys. Rev. B \textbf{83}, 104423 (2011).


\bibitem{Nijs}
 M. den Nijs and K. Rommelse, Phys. Rev. B \textbf{40}, 4709 (1989).

\bibitem{Tasaki}
 H. Tasaki, Phys. Rev. Lett. \textbf{66}, 798 (1991).




\bibitem{Haldane}
 F. D. M. Haldane, Phys. Lett. A \textbf{93}, 464 (1983).



%
 \bibitem{Vidal}
 G. Vidal, Phys. Rev. Lett. \textbf{91}, 147902 (2003);
 G. Vidal, Phys. Rev. Lett. \textbf{98}, 070201 (2007).



 \bibitem{Su}
 Y. H. Su, S. Y. Cho, B. Li, H. L. Wang, and H.-Q. Zhou,
 J. Phys. Soc. Jpn. \textbf{81}, 074003 (2012).


%
 \bibitem{Cardy}
 P. Calabrese and J. Cardy, J. Phys. A: Math. Theor. \textbf{42}, 504005 (2009);
 J. Cardy, {\it Scaling and Renormalization in Statistical Physics},
 (Oxford, University of Oxford, 1996).




 \bibitem{Tagliacozzo}
 L. Tagliacozzo, T. R. de Oliveira, S. Iblisdir and J. I. Latorre,
 Phys. Rev. B \textbf{78}, 024410(2008);
 F. Pollmann, S. Mukerjee, A. Turner and J. E. Moore,
 Phys. Rev. Lett. \textbf{102}, 255701 (2009);
 G. Vidal, J. I. Latorre, E. Rico and A. Kitaev,
 Phys. Rev. Lett. \textbf{90}, 227902 (2003).



 %
 \bibitem{Zhao}
 J.-H. Zhao, H.-L. Wang, B. Li, and H.-Q. Zhou, Phys. Rev. E \textbf{82}, 061127
 (2010).

%
 \bibitem{Su2}
 Y. H. Su, B.-Q. Hu, S.-H. Li, and S. Y. Cho,
 Phys. Rev. E \textbf{88}, 032110 (2013).
%



\bibitem{Zhou08}
 H.-Q. Zhou and J.P. Barjaktarevi$\check{\rm c}$,
 J. Phys. A: Math. Theor. {\bf 41}, 412001(2008);
 H.-Q. Zhou, R. Or\'us, and G. Vidal,
 Phys. Rev. Lett. {\bf 100}, 080601 (2008);
 H.-Q. Zhou, J.-H. Zhao, and B. Li,
 J. Phys. A: Math. Theor. \textbf{41}, 492002 (2008).



\bibitem{Hida1} 
 K. Hida, Phys. Rev. B \textbf{45}, 2207 (1992);
%
 K. Hida, Phys. Rev. B \textbf{46}, 8268 (1992).

\bibitem{Hyman} 
 R. A. Hyman, Kun Yang, R. N. Bhatt, and S. M. Girvin,
 Phys. Rev. Lett. \textbf{46}, 839  (1996).





\bibitem{Oshikawa}
 M. Oshikawa, J. Phys. Condens. Matter \textbf{4}, 7469 (1992).

%
\bibitem{Kennedy}
 T. Kennedy and H. Tasaki, Phys. Rev. B \textbf{45}, 304 (1992).
%


%
 \bibitem{Chaikin}
 P. M. Chaikin and T. C. Lubensky,
 {\it Pinciples of Condensed Matter Physics} (Cambridge University, Cambridge, 1995)
 pp 172-188.
%

 \bibitem{Fisher}
 M. E. Fisher and D. R. Nelson,
 Phys. Rev. Lett. \textbf{32}, 1350 (1974).

%

 \bibitem{Rohrer}
 H. Rohrer and Ch. Gerber,
 Phys. Rev. Lett. \textbf{38}, 909 (1977).

 \bibitem{Luther}
 A. Luther and D. J. Scalapino,
 Phys. Rev. B \textbf{16}, 1153 (1977).

 \bibitem{Hu}
 S. Hu, B. Normand, X. Wang, and L. Yu,
 Phys. Rev. B \textbf{84}, 220402R (2011).

 \bibitem{Kohmoto}
 M. Kohmoto, M. den Nijs, and L. P. Kadanoff,
 Phys. Rev. B \textbf{24} 5229 (1981).


\bibitem{Rams}
 M. M. Rams and B. Damski,
 Phys. Rev. Lett. \textbf{106}, 055701 (2011).


%
\end{thebibliography}
\end{document}